\documentclass[aps,prb,nobibnotes,twocolumn,superscriptaddress,bibliography]{revtex4-2}
\usepackage{amsfonts}
\usepackage{mathrsfs}
\usepackage{amsmath}
\usepackage{color}
\usepackage{graphicx}
\usepackage{bm}
\usepackage{amssymb}
\usepackage{xspace}
\usepackage{epstopdf}
\usepackage{dcolumn}
\usepackage{longtable}
\usepackage{multirow}
\usepackage{float}
\usepackage{comment}

\usepackage[colorlinks=true, letterpaper=true, pdfstartview=FitV, linkcolor=blue, citecolor=blue, urlcolor=blue]{hyperref}

\makeatother

\begin{document}
\title{Anomalous shift in Andreev reflection from side incidence }
\author{Runze Li}
\affiliation{Centre for Quantum Physics, Key Laboratory of Advanced Optoelectronic
Quantum Architecture and Measurement (MOE), School of Physics, Beijing
Institute of Technology, Beijing, 100081, China}
\affiliation{Beijing Key Lab of Nanophotonics \& Ultrafine Optoelectronic Systems,
School of Physics, Beijing Institute of Technology, Beijing, 100081,
China}

\author{Chaoxi Cui}
\affiliation{Centre for Quantum Physics, Key Laboratory of Advanced Optoelectronic
Quantum Architecture and Measurement (MOE), School of Physics, Beijing
Institute of Technology, Beijing, 100081, China}
\affiliation{Beijing Key Lab of Nanophotonics \& Ultrafine Optoelectronic Systems,
School of Physics, Beijing Institute of Technology, Beijing, 100081,
China}

\author{Ying Liu}
\email{ying\_liu@hebut.edu.cn}
\affiliation{School of Materials Science and Engineering, Hebei University of Technology, Tianjin 300130, China}

\author{Zhi-Ming Yu}
\email{zhiming\_yu@bit.edu.cn}

\affiliation{Centre for Quantum Physics, Key Laboratory of Advanced Optoelectronic
Quantum Architecture and Measurement (MOE), School of Physics, Beijing
Institute of Technology, Beijing, 100081, China}
\affiliation{Beijing Key Lab of Nanophotonics \& Ultrafine Optoelectronic Systems,
School of Physics, Beijing Institute of Technology, Beijing, 100081,
China}

\author{Shengyuan A. Yang}
\affiliation{Research Laboratory for Quantum Materials, IAPME, Faculty of Science and Technology, University of Macau, Macau SAR}

\begin{abstract}
Andreev reflection at a normal-superconductor interface may be accompanied with an anomalous spatial shift. The studies so far are limited to the top incidence configuration. Here, we investigate this effect in the side incidence configuration, with the interface parallel to the principal axis of superconductor. We find that the shift exhibits rich behaviors reflecting the character of pair potential. It has two contributions: one from the $k$-dependent phase of pair potential, and the other from the evanescent mode. For chiral $p$-wave pairing, the pairing phase contribution is proportional to the chirality of pairing and is independent of excitation energy, whereas the evanescent mode contribution is independent of chirality and is nonzero only for excitation energy below the gap. The two contributions also have opposite parity with respect to the incident angle. For $d_{x^{2}-y^{2}}$-wave pairing, only the
evanescent mode contribution exists, and the shift exhibits suppressed zones in incident angles, manifesting the superconducting nodes. The dependence of the shift on other factors, such as the angle of incident plane and Fermi surface anisotropy, are discussed.

\end{abstract}
\maketitle

\section{INTRODUCTION \label{sec:I}}
Many interesting optical effects have found their analogies in electronic systems.
In geometric optics, a well-known
phenomenon is the anomalous spatial shift of a light beam during reflection at an optical interface~\cite{bliokh2013,bliokh2015,ling2017}. With reference to the beam's incident plane, this shift may be decomposed into two components: the longitudinal component which is within the plane, known as the Goos-H\"{a}nchen effect~\cite{Goos1947}; and the transverse component which is normal to the plane, known as the Imbert-Fedorov
effect~\cite{IF1955,IF1972}. These effects have been extensively studied in both theory and experiment, and they have found wide applications in interface characterization, biological sensing, nanophotonics, and etc~\cite{bliokh2013,bliokh2015,ling2017,Chen2023}.

The analogous effects for electrons, namely, the shifts for an electron beam when scattered at an interface, also exist.
The longitudinal (Goos-H\"{a}nchen-like) shift was studied already in the 1970s~\cite{SCM1972,DMF1974,DMF21974}. With the technological advance which makes it possible to achieve precise control of electron beam trajectory (which leads to the field of electron optics)~\cite{EO1,EO2,EO3},
these electronic shifts have attracted increasing interest in the past two decades. Notably, it was found that the shifts often encode key features of electronic band structures of materials that form the interface. For example, the Goos-H\"{a}nchen-like shift in graphene has strong dependence on the Dirac electron's pseudospin degree of freedom~\cite{Beenakker2009,chen2011,sharma2011,wu2011}; and the Imbert-Fedorov-like shift first reported for interface with Weyl semimetals is sensitive to the chirality of Weyl points~\cite{jiang2015,Yang2015,jiang2016,wang2017,chatto2019}.
More recently, it was shown that under certain symmetry conditions, the shift could lead to a quantized circulation pattern when plotted in momentum space, and this may capture topological characters of a bulk material~\cite{Yliu2020}.

In optic and electronic cases mentioned above, the shift occurs in a reflection in which the incident and the reflected beams are of the same kind of particles. However, at the interface between a normal metal and a superconductor, there is a unique reflection process, the Andreev reflection, where an incident electron is reflected back as a hole and vice versa.
Although the particle identity is changed during Andreev reflection, in 2017, Liu \emph{et al.}~\citep{Yliu2017} showed that
a spatial shift still exists in this process. Interestingly, the shift is sensitive to the type of superconducting pairing~\citep{Yu2018}.
For example, consider the setup where an electron beam is incident from a simple medium (e.g., vacuum) and hits the interface with a superconductor. It was found that $s$-wave pairing leads only to longitudinal shift~\citep{1Yliu2018}. Yu \emph{et al.}~\citep{Yu2018} showed that, in comparison, for unconventional pairings, such as $d$-wave or chiral $p$-wave pairings, both longitudinal and transverse shifts occur and they manifest intriguing features unique for each pairing symmetry. Thus,
the effect of spatial shift in Andreev reflection provides a powerful tool for characterizing superconductivity.

In previous studies, the normal-superconductor (NS) interface is taken to be the one normal to the $c$-axis (in other words, parallel to the $ab$-plane) of the superconductor, a setup which may be called the top incidence configuration.  It is noted that unconventional pairings, like $p$-wave and $d$-wave pairings, are anisotropic~\citep{sigrist1991,tsuei2000,bergeret2005}. This indicates that the physics could be very different for the scattering happening on the side surface parallel to the $c$-axis. We call this setup the side incidence configuration. Then, a natural question is: Does the shift exist also in the side incidence configuration? If yes, what are its special features, particularly in comparison with the top incidence configuration?


In this work, we investigate the anomalous spatial shift in Andreev reflection in the side incidence configuration and answer the above questions. Specifically, we consider two unconventional pairing models, with chiral $p$-wave pairing and $d_{x^{2}-y^{2}}$-wave pairing, respectively. We show that the shift generally exists, can be sizable, and exhibits features distinct from top incidence.
For the chiral $p$-wave case, the behavior of the shift becomes particularly simple for excitation energy $\varepsilon$ above the pairing gap, where
both longitudinal and transverse shifts become independent of $\varepsilon$, and their signs are determined by the chirality of pairing. Meanwhile, for $\varepsilon$ below the pairing gap, there is an additional contribution to each shift component, which is not related to chirality. As a result, while the shift is symmetric in the incident angle for high excitation energy (above the gap), it is no longer symmetric for low excitation energy (below the gap).
For the $d_{x^{2}-y^{2}}$-wave pairing, there exist zones of incident angle for nonzero excitation energy $\varepsilon$ where the shift is completely suppressed, which correspond to the nodes of the pairing gap. Both longitudinal and transverse shifts are enhanced when $\varepsilon$ is close but below the pairing gap seen by the incident electron.
Our work clarifies the intriguing effect of spatial shift in Andreev reflection in an important setup. The result complements the previous studies on top incidence to provide a complete picture, which deepens our understanding of this fundamental effect and can be useful for superconductor characterization as well as device design.

\section{Basic setup and Modelling} \label{sec:II}

\begin{figure}[t]
\includegraphics[width=8.7cm]{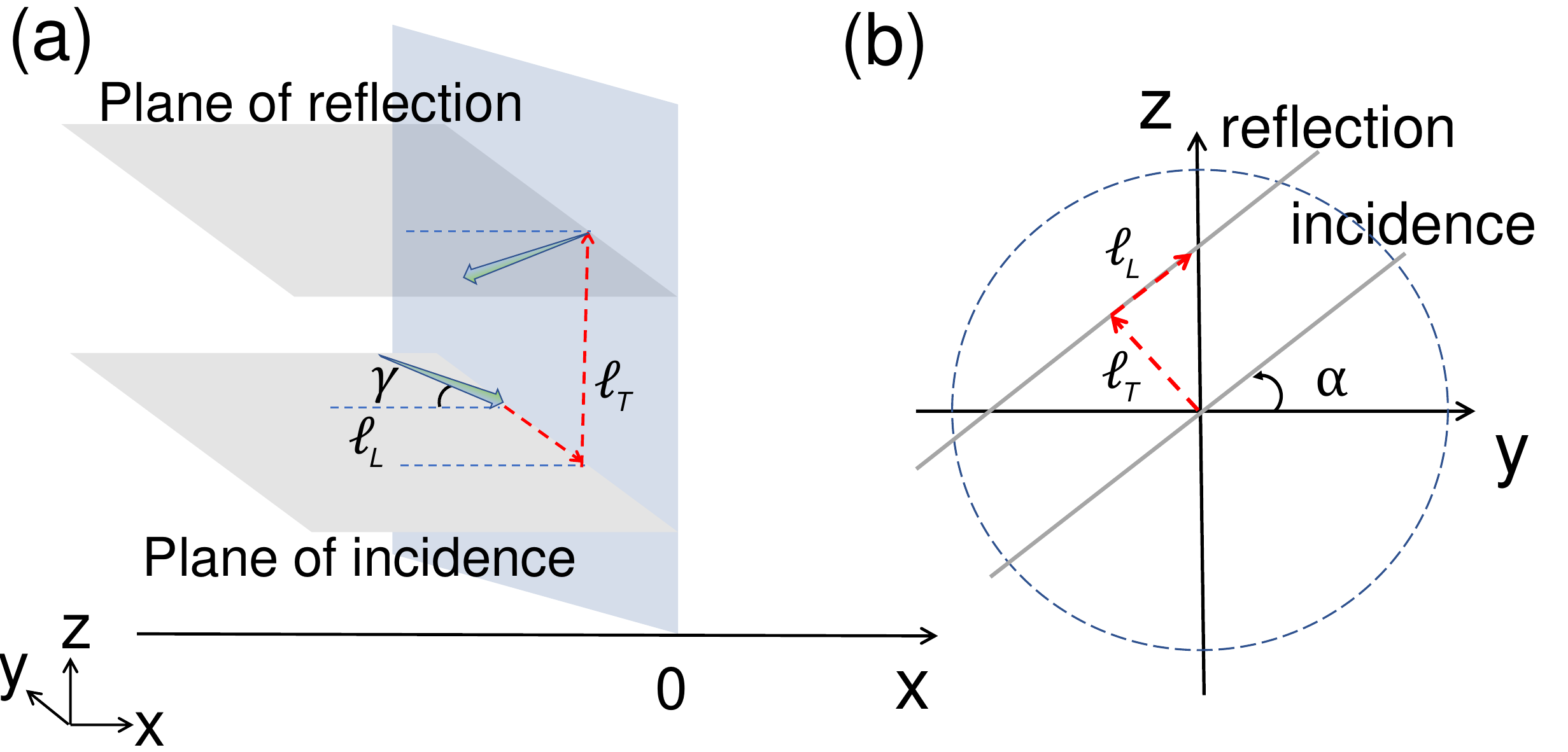} \caption{(a) Schematic figure showing the basic setup for investigating the
longitudinal and transverse shifts in interface scattering. The green arrows denote the incident and the reflected particle beams. (b) The two shift components $\ell_L$ and $\ell_T$ viewed in $y$-$z$ plane. \label{fig: illustrations}}
\end{figure}

Let us first describe the basic setup of side incidence configuration. As illustrated in Fig.~\ref{fig: illustrations}(a), we consider a flat NS interface. The left side ($x<0$) is occupied by the normal (N) medium (e.g., simple metal or vacuum). The right side ($x>0$) is occupied by the superconducting (S) medium. The system is extended in the $y$ and $z$ directions, so momenta $k_y$ and $k_z$ are conserved during scattering at the interface. Consider an electron beam incident from the N side.
The incident plane is defined by the beam and the interface normal vector. We use the angle
\begin{equation}
  \alpha=\tan^{-1}(k_{z}/k_{y})\in (-\pi,\pi]
\end{equation}
to specify the incident plane. As shown in Fig.~\ref{fig: illustrations}(b), $\alpha$ is the angle of incident plane measured from the $y$ axis. Inside the incident plane, the incident angle of the beam is denoted by $\gamma\in (-\pi/2,\pi/2)$ [see Fig.~\ref{fig: illustrations}(a)], and $\gamma=0$ corresponds to the case of normal incidence.


The essential physics of scattering at the NS interface can be described by the Bogoliubov-de Gennes (BdG) equation \citep{Blonder1982,de1999}:
\begin{equation}
\left[\begin{array}{cc}
H_{0}+V(x)-E_{F} & \boldsymbol\Delta^{*}(\boldsymbol{k})\Theta(x)\\
\boldsymbol\Delta(\boldsymbol{k})\Theta(x) & E_{F}-H_{0}-V(x)
\end{array}\right]\psi=\varepsilon\psi,\label{ham}
\end{equation}
where $H_{0}=-\frac{1}{2m}\nabla^2$ is the kinetic energy term (we set $\hbar=e=1$), $E_{F}$ is the Fermi energy, $\boldsymbol\Delta(\boldsymbol{k})$ is the superconducting pair potential,
$\Theta(x)$ is the Heaviside step function indicating that the pairing occurs on the S side,
and $V(x)=U\Theta(x)+h\delta(x)$
with $U$ denoting a potential energy difference between the two sides and $h$ denoting a possible interface barrier potential. For side incidence configuration, the principal axis of the S side is along $z$. For unconventional pairings, the pair potential  $\boldsymbol\Delta$ should have strong dependence on $k_x$ and $k_y$, and we will neglect its $k_z$ dependence. For example, for the chiral $p$-wave pairing, we write $\boldsymbol\Delta=\Delta_0(k_x+i\chi k_y)$, where parameters $\Delta_0>0$ and $\chi=\pm 1$. In this model, we take isotropic Fermi surfaces. In practice, unconventional superconductors often have anisotropic Fermi surfaces. We will discuss effects of anisotropic Fermi surface later in Sec.~\ref{sec:VII}.

%
%

\section{analytic results}\label{sec:III}
\begin{figure}[t]
\includegraphics[width=8.7cm]{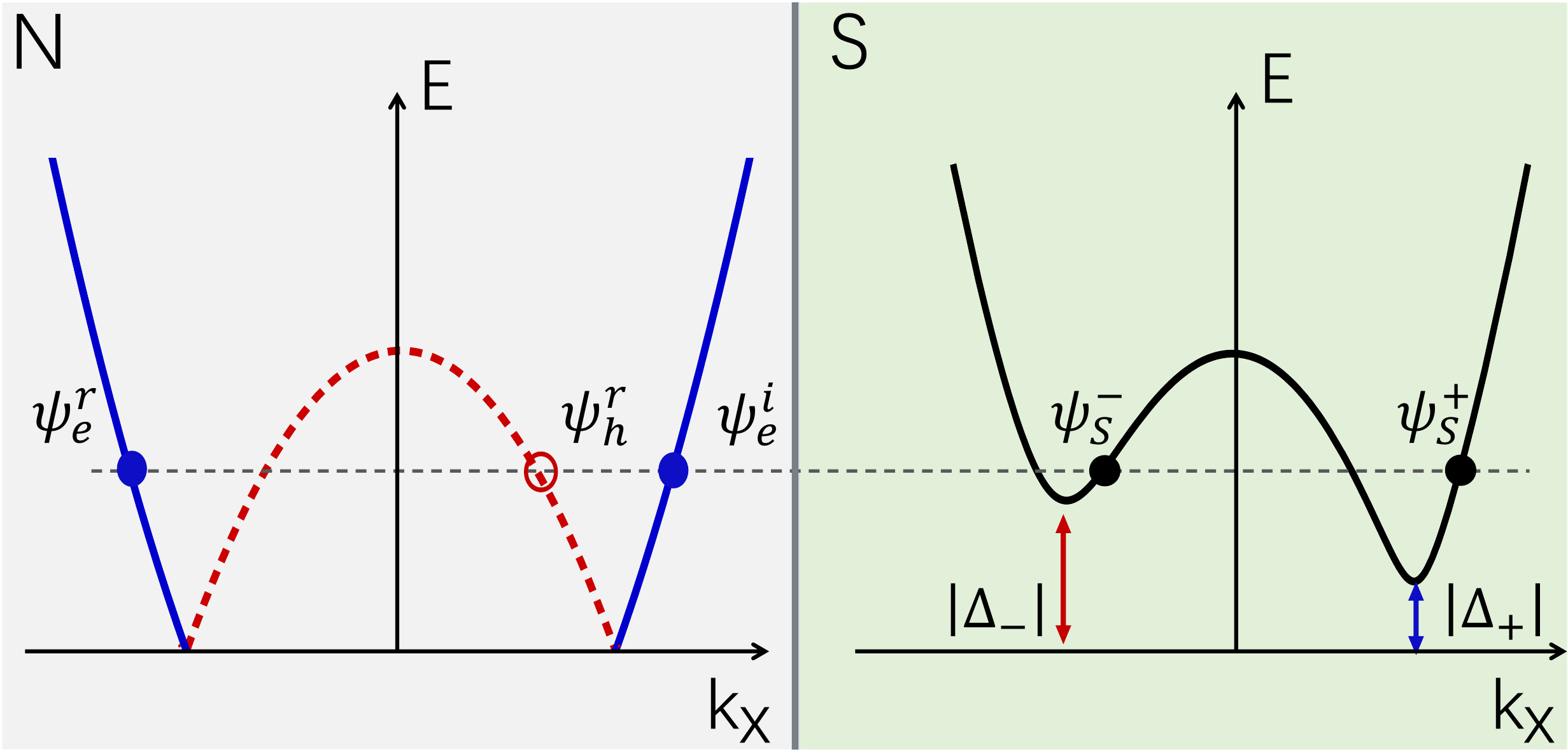} \caption{Schematic figure showing the states involved in the scattering process. The blue and (hollow) red dots denotes the electron and hole states in N region. The black dots denote the two quasiparticle states in S region. The superconducting gaps $\Delta_\pm$ around $\psi_S^\pm$ states may have different values in the side incidence configuration. \label{fig: ill-bands}}
\end{figure}

Before performing calculations, we first analyze states that are involved in a scattering. Since the momentum $\bm k_\|=(k_y,k_z)$ parallel to the interface is conserved, in our model, there will be five states involved (see Fig.~\ref{fig: ill-bands}): incident electron state $\psi_e^i$, reflected electron state $\psi_e^r$, reflected hole state $\psi_h^r$, and two transmitted states $\psi_S^+$ and $\psi_S^-$. In Fig.~\ref{fig: ill-bands}, one can see that $\psi_S^+$ is an electron-like quasiparticle state, whereas $\psi_S^-$ is a hole-like quasiparticle state. It should be noted that the superconducting gaps around the locations of $\psi_S^+$ and $\psi_S^-$ could be different, due to the $k$ dependence of $\boldsymbol\Delta$ for unconventional pair potential. We denote the two gaps as $\Delta_+$ and $\Delta_-$, respectively (see Fig.~\ref{fig: ill-bands}). They generally depend on the energy and momentum of the incident electron. We define their phase angles as
\begin{equation}\label{theta}
  \theta_\pm=\arg(\Delta_{\pm}).
\end{equation}
For the case illustrated in Fig.~\ref{fig: ill-bands}, $\psi_S^+$ and $\psi_S^-$ are propagating modes in S. When the excitation energy $\varepsilon$ is below $\Delta_+$ ($\Delta_-$), $\psi_S^+$ ($\psi_S^-$) will become an evanescent mode.


Now, for a given incident electron state $\psi_e^i$, we can write down the corresponding scattering state for the BdG equation (\ref{ham}):
\begin{equation}
\psi=\begin{cases}
\psi_{e}^{i}+r_{e}\psi_{e}^{r}+r_{h}\psi_{h}^{r}, & x<0\\
t_{+}\psi_{S}^{+}+t_{-}\psi_{S}^{-}, & x>0
\end{cases},
\label{Sfunction}
\end{equation}
where $r_{e(h)}$ is the amplitude for normal (Andreev) reflection, and $t_\pm$ are the two amplitudes for transmissions into $\psi_S^\pm$ states.

The three states in the N region can be written as
\begin{eqnarray}
\psi_{e}^{i}&=&e^{+ik_{e}x}\left(\begin{array}{l}
1\\
0
\end{array}\right),\\
\psi_{e}^{r}&=&e^{-ik_{e}x}\left(\begin{array}{l}
1\\
0
\end{array}\right),\\
\psi_{h}^{r}&=&e^{+ik_{h}x}\left(\begin{array}{l}
0\\
1
\end{array}\right),
\end{eqnarray}
with $k_{e}, k_{h}\approx\sqrt{2mE_F-k_{\|}^{2}}$ and $k_{\|}=\text{sgn}(\gamma)\sqrt{k_y^2+k_z^2}$. Here, we assume $E_F\gg \varepsilon$, so corrections to $k_{e/h}$ of order $(\varepsilon/E_F)$ are neglected. From the configuration in Fig.~\ref{fig: illustrations}, we have the relation $\tan\gamma=k_\|/k_e$.

In the S region, the  basis states $\psi_{S}^{\pm}$ (not normalized) can be written as
\begin{equation}
\psi_{S}^{\pm}=\left(\begin{array}{c}
1\\
\eta_{\pm}e^{i\theta_{\pm}}
\end{array}\right)e^{ik_{S\pm}x}.\label{SCstate}
\end{equation}
Here, $\theta_{\pm}$ have been defined in Eq.~(\ref{theta}) which are phases of pair potential, and
\begin{equation}
  \eta_{\pm}=\frac{|\Delta_{\pm}|}{\varepsilon\pm\sqrt{\varepsilon^2-|\Delta_{\pm}|^2}}
\end{equation}
and the momentum $k_{S\pm}\approx \pm k_S$, with
\begin{equation}
k_{S} = \sqrt{2m(E_F-U)-k_{\|}^2}.
\end{equation}
In the treatment, we take the weak coupling limit, with $(E_{F}-U)\gg \{|\boldsymbol\Delta|,\varepsilon\}$. Meanwhile, the magnitudes of
$|\boldsymbol\Delta|$ and $\varepsilon$ can be comparable. One notes that for $\varepsilon>|\Delta_{\pm}|$, $\eta_\pm$ is a positive (real) number, whereas for  $\varepsilon<|\Delta_{\pm}|$, $\eta_\pm$ becomes complex.



The boundary conditions for the BdG equation at the interface $(x=0)$ are derived from the quasiparticle current conservation~\citep{Blonder1982,Ka2000}. They take the form of
\begin{equation}
\begin{aligned}
\left.\psi\right|_{x=0^{-}} & =\left.\psi\right|_{x=0^{+}}, \\
\left.\frac{1}{m} \partial_x \psi\right|_{x=0^{-}} & =\left.\frac{1}{m} \partial_x \psi\right|_{x=0^{+}}-2 h \psi(0).
\end{aligned}
\label{BC}
\end{equation}

The scattering amplitudes ($r$'s and $t$'s) can be solved from  Eq.~(\ref{BC}) by substituting Eq.~(\ref{Sfunction}).
The information of the spatial shift in Andreev reflection is encoded in the scattering amplitude $r_h$. After straightforward calculations, we obtain
\begin{equation}
r_{h}=-\frac{4\lambda\eta_{+}\eta_{-}e^{i\left(\theta_{+}+\theta_{-}\right)}}{\eta_{+}\Lambda_{-}e^{i\theta_{+}}
-\eta_{-}\Lambda_{+}e^{i\theta_{-}}},\label{amp}
\end{equation}
where we define dimensionless parameters $\lambda={k_{S}}/{k_{e}}$ and
\begin{equation}
  \Lambda_\pm=4(mh/k_e)^2+(1\pm \lambda)^2.
\end{equation}

Based on the results in Ref.~\cite{Yliu2017,Yu2019},
the spatial shift $\bm \ell$ for the Andreev reflected hole beam can be calculated from
\begin{equation}
\ell_{i}=-\left.\frac{\partial}{\partial k_{i}}\varphi\right|_{\boldsymbol{k}_{\parallel}}\qquad (i=y,z),\label{shift}
\end{equation}
with $\varphi=\arg (r_h)$. The shift depends on the phase of $r_h$, not its magnitude. It is customary to decompose the shift vector $\bm \ell$ into the longitudinal component $\ell_L$ inside the incident plane and the transverse component $\ell_T$ normal to the plane (see Fig.~\ref{fig: illustrations}). In the present setup, we have
\begin{equation}
  \ell_L=\ell_y\cos\alpha + \ell_z \sin\alpha,
\end{equation}
and
\begin{equation}
  \ell_T=-\ell_y\sin\alpha + \ell_z \cos\alpha.
\end{equation}
\\

Now, to obtain the phase angle $\varphi$, by using the result in  Eq.~(\ref{amp}), we find
\begin{equation}
\tan\varphi=\begin{cases}
\frac{\Lambda_{+}\sin(\phi_{+}+\theta_{+})-\Lambda_{-}\sin(\phi_{-}+\theta_{-})}
{\Lambda_{+}\cos(\phi_{+}+\theta_{+})-\Lambda_{-}\cos(\phi_{-}+\theta_{-})}, & \varepsilon<|\Delta_{\pm}|\\
\\
\frac{\eta_{-}\Lambda_{+}\sin\theta_{+}-\eta_{+}\Lambda_{-}\sin\theta_{-}}{\eta_{-}\Lambda_{+}\cos\theta_{+}-\eta_{+}
\Lambda_{-}\cos\theta_{-}}, & \varepsilon>|\Delta_{\pm}|
\end{cases}\label{arg1}
\end{equation}
where
\begin{equation}
  \phi_{\pm}=\arg(\eta_{\pm}).
\end{equation} Since $\Delta_+$ and $\Delta_-$ may have different values, besides the two cases in Eq.~(\ref{arg1}), we have another two cases:
when $|\Delta_{-}|<\varepsilon<|\Delta_{+}|$, we find
\begin{equation}
\tan\varphi=\frac{\Lambda_{-}\sin(\theta_{-})-\eta_{-}\Lambda_{+}\sin(\phi_{+}+\theta_{+})}{\Lambda_{-}\cos(\theta_{-})-
\eta_{-}\Lambda_{+}\cos(\phi_{+}+\theta_{+})}; \label{arg1-1}
\end{equation}
and when $|\Delta_+|<\varepsilon<|\Delta_{-}|$,
\begin{equation}
\tan\varphi=\frac{\Lambda_{+}\sin(\theta_{+})-\eta_{+}\Lambda_{-}\sin(\phi_{-}+\theta_{-})}
{\Lambda_{+}\cos(\theta_{+})-\eta_{+}\Lambda_{-}\cos(\phi_{-}+\theta_{-})}.
\label{arg1-2}
\end{equation}
\\

Before proceeding, let's make a comparison with  top incidence configuration. In that case, one always has the equality $\Delta_+=\Delta_-$, so essentially the subscripts $\pm$ can be dropped in quantities involved in the formulas above. This will greatly simplify the result. One can deduce that
%
\begin{equation}
\varphi=\begin{cases}
\theta+\tan^{-1}(\zeta\tan\phi), & \varepsilon<|\Delta_+|,\\
\theta, & \varepsilon>|\Delta_+|,
\end{cases}\label{arg2}
\end{equation}
where we define $\zeta=[{4(mh/k_e)^{2}+1+\lambda^{2}}]/{(2\lambda)}$, $\theta=\theta_{\pm}$, and $\phi=\phi_{\pm}$. This recovers the previous result in Ref.~\cite{Yu2018}.

Back to side incidence configuration considered here, generally, we have $\Delta_{+}\neq \Delta_{-}$. Then, the resulting expressions of $\varphi$, i.e., Eqs.~(\ref{arg1}-\ref{arg1-2}), are more complicated.
Nevertheless, if we restrict to the regime where $E_F$ is the largest energy scale, i.e., with $E_F\gg \{U, h\}$,
and $|\gamma|$ not close to $\pi/2$,
then we have {$k_e\approx k_S$},
$\lambda\approx 1$, $\Lambda_+\approx 4$, and $\Lambda_-\approx 0$,
so that
\begin{equation}\label{22}
  r_h\approx \eta_+ e^{i\theta_+}.
\end{equation}
In this regime, the expression for $\varphi$ is simplified to
%
\begin{equation}
\varphi=\begin{cases}
\theta_{+}+\phi_{+}, & \varepsilon<|\Delta_{+}|,\\
\theta_{+}, & \varepsilon>|\Delta_{+}|.
\end{cases}\label{arg3}
\end{equation}

This is a very nice result. We have the following observations. First, the result depends on the $\psi_S^+$ state but not
$\psi_S^-$ state. This can be intuitively understood, because for large $E_F$, $\psi_S^-$ is largely separated in momentum from $\psi_e^i$, $\psi_S^+$, and $\psi_h^r$ states, so it has little influence on Andreev reflection. Second, in Eq.~(\ref{arg3}), $\varphi$ has two contributions when $\varepsilon$ is below the superconducting gap.   The first contribution $\theta_+$ originates from the phase of pair potential.
The second contribution $\phi_+$ originates from the evanescent character of $\psi_S^+$ state in this case. Indeed, previous studies showed that evanescent modes play a critical role in generating the spatial shift.
These are the two sources of the phase change between the incident electron and the Andreev reflected hole.
Third, when the excitation energy is above the gap, we only have the $\theta_+$ contribution. This is because $\psi_S^+$ state now becomes a propagating mode, which then does not contribute a phase change.

In the following sections, we will apply the above formulas to three different types of pair potentials on the S side.



\section{s-wave pairing}\label{sec:IV}

Let's first apply the results in Sec.~\ref{sec:III} to the case of conventional $s$-wave pair potential. Since this case is isotropic,
the result for side incidence should be the same as that for top incidence. The purpose of the discussion here is mainly for completeness and also to provide a reference to which the results of unconventional pairing cases can be compared.

\begin{figure}[t]
\includegraphics[width=8.7cm]{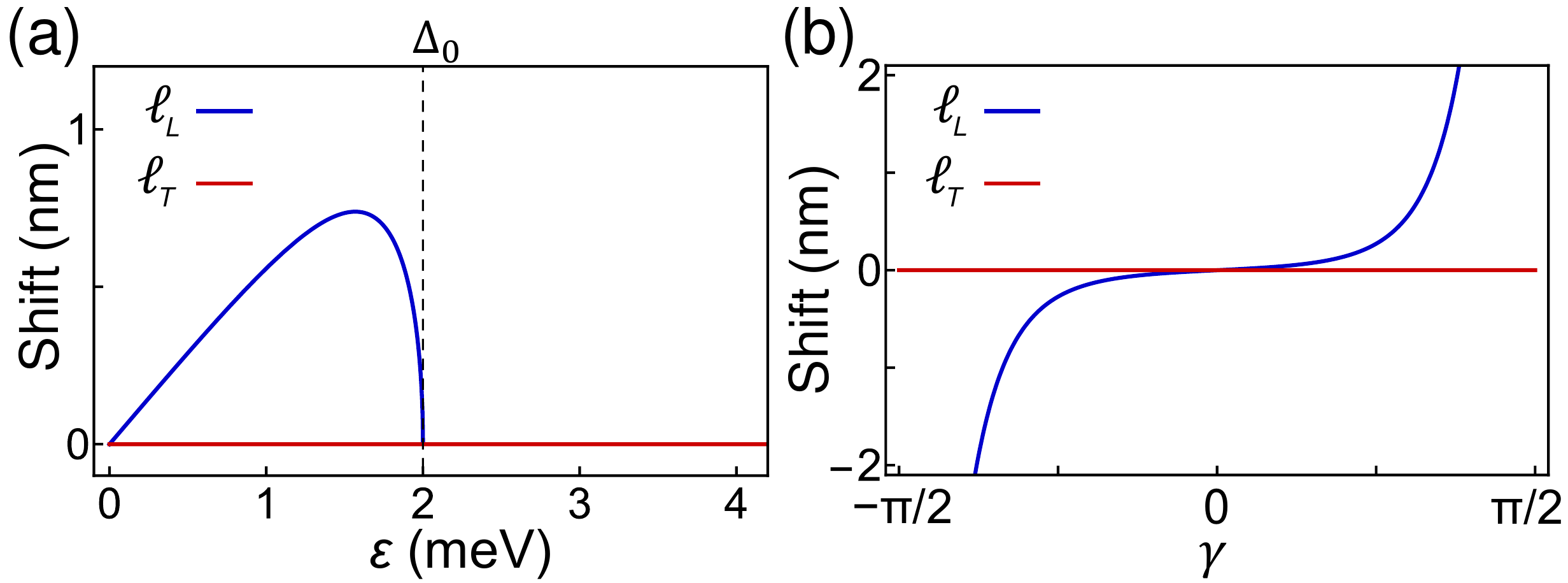} \caption{Results for the shifts in Andreev reflection for the $s$-wave case. (a) shows spatial shifts versus $\varepsilon$. (b) shows the spatial shifts versus the incident angle $\gamma$. In the calculation, we set $E_F=0.2\ {\rm {eV}}$, $U=-0.2\ {\rm {eV}}$, $m=0.1\  m_e$, $h=50\ \rm{meV}\cdot\rm{nm}$ and $\Delta_0=2\ {\rm {meV}}$; in (a), we take $\gamma=0.3\pi$; in (b), we take  $\varepsilon=1\ {\rm {meV}}$.\label{figswave}}
\end{figure}

For a $s$-wave pair potential, we model $\boldsymbol\Delta=\Delta_0$ being a real constant parameter. By using the result in Sec. \ref{sec:III}, especially the analysis around Eq.~(\ref{arg2}),
{we find that the two components of the shift are given by
 \begin{equation}
   \ell_{T} = 0,
 \end{equation}
\begin{equation}\label{slL}
  \ell_{L} = -\frac{8 m^2 \tan \phi (U^2+h^2 k_N^2)  k_{\|}}{4k_e^3 k_S^{3} +k_e k_S(k_N^2+4  m^2 h^2)^2 \tan^2\phi},
\end{equation}
for $\varepsilon<\Delta_0$; and
\begin{equation}
  \ell_T=\ell_L=0,
\end{equation} for $\varepsilon>\Delta_0$.
In the above expression, $k_N=\sqrt{k_e^2+k_S^{2}}$ and $\phi=\phi_\pm=-\arccos \frac{\varepsilon}{\Delta_0}$.

In this case, the pair potential does not have a nontrivial phase variation, so a nonzero shift has to come from the evanescent mode contribution, which requires $\varepsilon<\Delta_0$.
Clearly, the result should not depend on angle $\alpha$, due to isotropy of the model.
One observes that the transverse component $\ell_T$ of the shift vanishes. This can be readily understood by noting that the system always has a mirror symmetry with respect to the incident plane. Regarding the longitudinal shift $\ell_L$,
it is an odd function of $\gamma$ [see Fig.~\ref{figswave}(b)], as $k_\|$ is odd in $\gamma$. From Eq.~(\ref{slL}), a finite $\ell_L$ would require a finite $U$ or $h$. We also note that
a large $E_F$ which dominates over $U$ and $\ell$ would suppress the value of $\ell_L$. This is because for such case, according to the discussion around Eq.~(\ref{22}), $\varphi=\phi_+=\phi$ becomes a $k$-independent number ($\theta_+=0$ for $s$-wave). Then the shift from Eq.~(\ref{shift}) should vanish. These results are consistent with the previous studies for $s$-wave case in the top incidence configuration~\cite{Yliu2017,Yu2018}.

\begin{figure}[t]
\includegraphics[width=8.7cm]{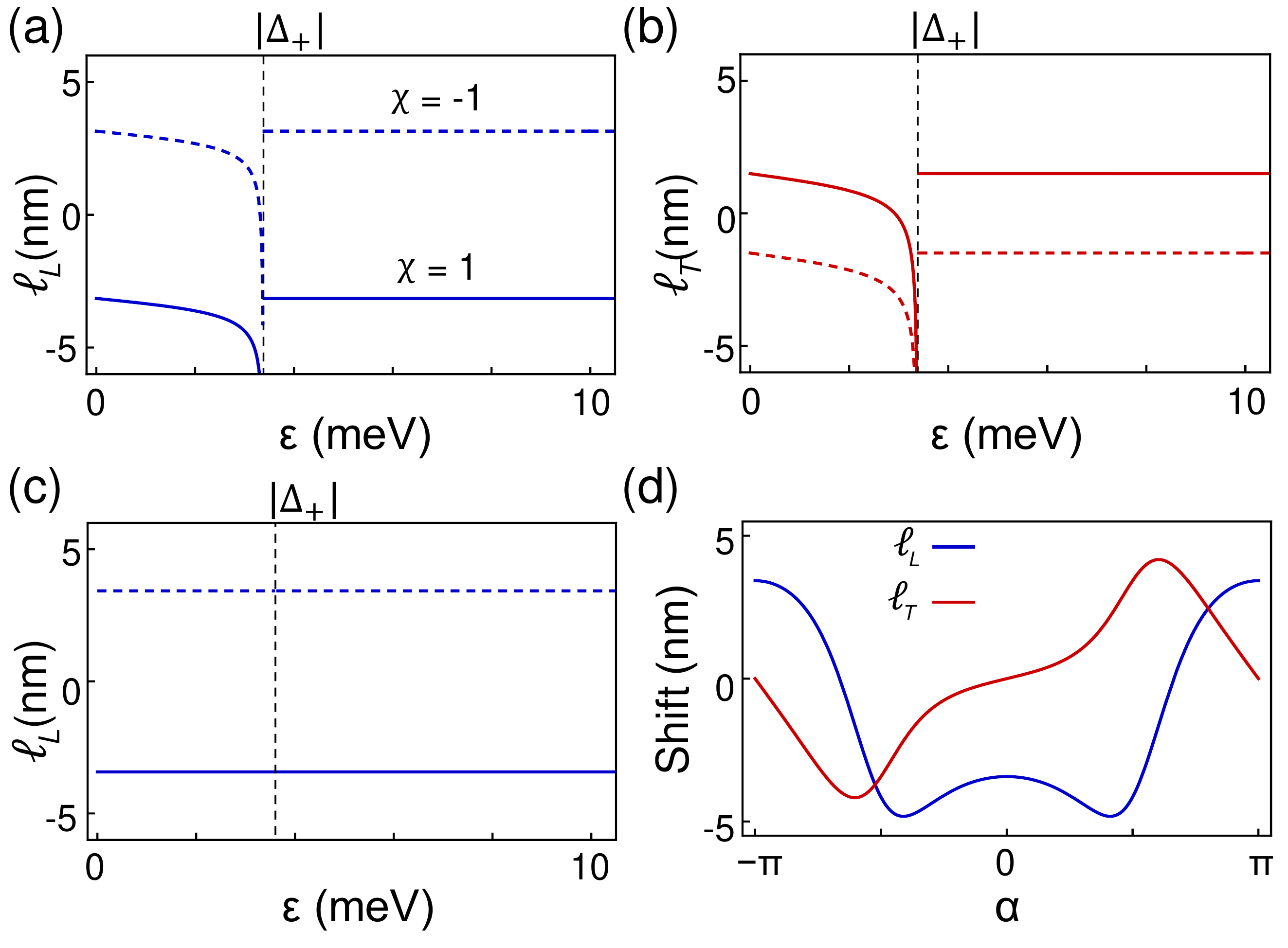} \caption{Calculated (a) longitudinal shift $\ell_L$ and (b) transverse shift $\ell_T$ as functions of excitation energy $\varepsilon$. In these figures, the solid (dashed) lines are results for chirality $\chi=+1$ ($-1$). (c) shows longitudinal shifts for the special configuration with angle $\alpha=0$. (d) denotes spatial shifts as functions of angle $\alpha$.
In the calculation, we set $E_F=0.1\ \rm{eV} $, $U=0$, $m=0.05\  m_e$, $\Delta_{0}=10\ \rm{meV}\cdot\rm{nm}$, $h=0$ and $\gamma=0.2\pi$. In (a) and (b), we take $\alpha=0.2\pi$; in (c), we take $\alpha=0\pi$; in (d), we take $\varepsilon=2.5\ \rm{meV}$;
\label{figpwave1}}
\end{figure}

\begin{figure}[t]
\includegraphics[width=8.7cm]{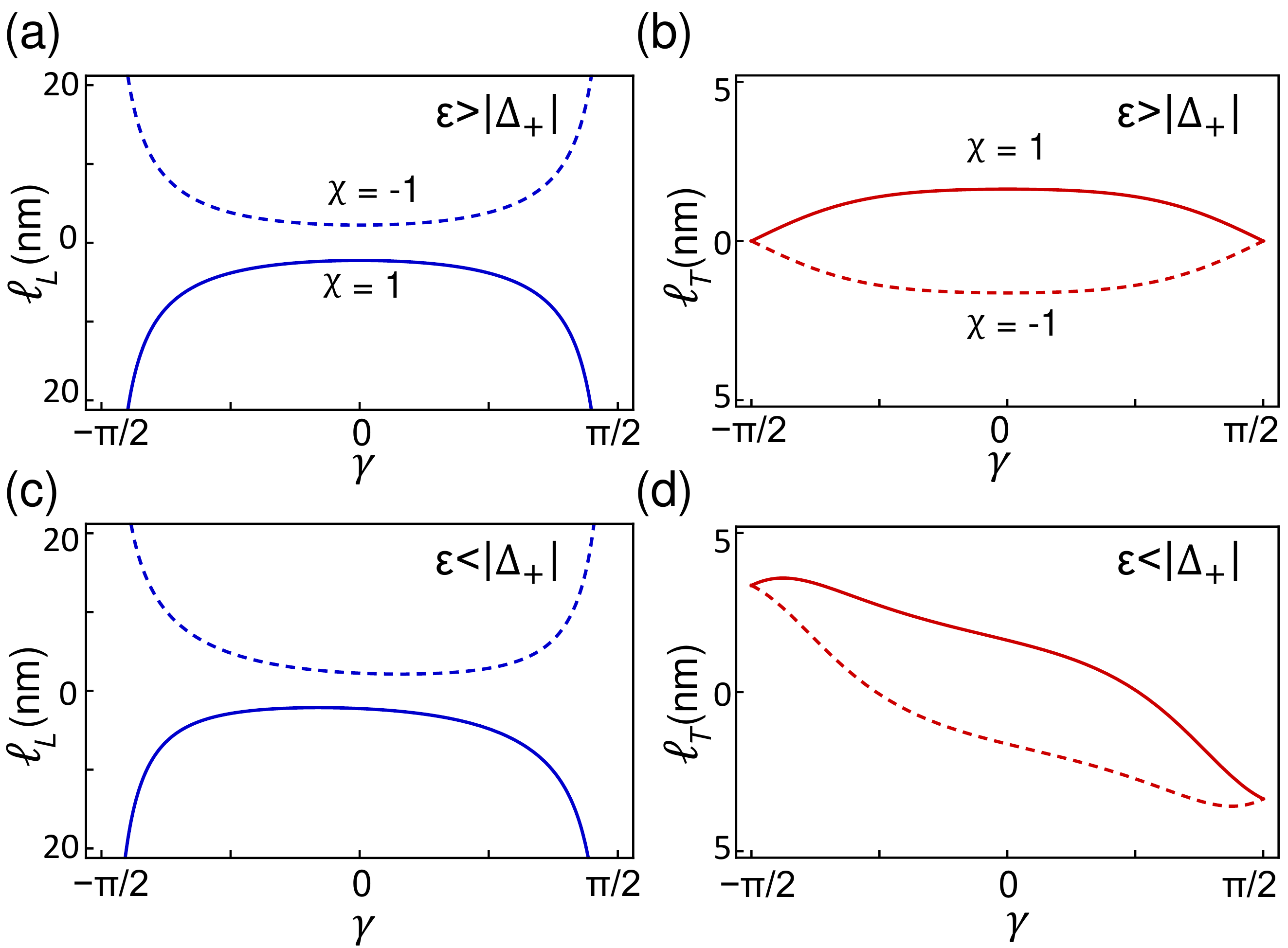} \caption{Calculated (a,c) longitudinal shift $\ell_L$ and (b,d) transverse shift $\ell_T$ as functions of incident angle $\gamma$. In these figures, the solid (dashed) lines are results for chirality $\chi=+1$ ($-1$). In the calculation, we set $E_F=0.1\ \rm{eV} $, $U=0$, $m=0.05\  m_e$, $\Delta_{0}=10\ \rm{meV}\cdot\rm{nm}$, $h=0$ and $\alpha=0.2\pi$; In (a) and (b), we take $\varepsilon=4\ \rm{meV}$; in (c) and (d), we take $\varepsilon=2.5\ \rm{meV}$;
\label{figpwave2}}
\end{figure}


\section{chiral p-wave pairing}\label{sec:V}
Next, we consider the case with a chiral $p$ -wave superconductor.  The pair potential on S side is modeled as $\boldsymbol{\Delta}=\Delta_{0}(k_{x}+i\chi k_{y})$, where $\chi=\pm1$ denotes the chirality.
For such a chiral pair potential, the two gaps $|\Delta_\pm|$ are the same, regardless of the angles $\gamma$ and $\alpha$.
Here, we focus on the regime with $E_F\gg \{U, h\}$.
By using  Eq.~(\ref{arg3}), the expressions of  the anomalous shift can be obtained as
\begin{equation}
\ell_{L}=\begin{cases}
-\chi\frac{k_{\|}^2+k_{S}^{2}}{k_S k_{S}^{'2}}\cos\alpha-\frac{\varepsilon k_{\|}\sin^{2}\alpha}{k_{S}^{'2}\sqrt{\Delta_{0}^{2}k_{S}^{'2}-\varepsilon^{2}}}, & \varepsilon<|\Delta_+|\\
-\chi\frac{k_{\|}^2+k_{S}^{2}}{k_S k_{S}^{'2}}\cos\alpha, & \varepsilon>|\Delta_+|
\end{cases}\label{pGHshift}
\end{equation}
for longitudinal shift, and
\begin{equation}
\ell_{T}=\begin{cases}
\chi\frac{k_{S}}{k_{S}^{'2}}\sin\alpha-\frac{\varepsilon k_{\|}\sin2\alpha}{2k_{S}^{'2}\sqrt{\Delta_{0}^{2}k_{S}^{'2}-\varepsilon^{2}}}, & \varepsilon<|\Delta_+|\\
\chi\frac{k_{S}}{k_{S}^{'2}}\sin\alpha, & \varepsilon>|\Delta_+|
\end{cases}\label{pIFshift}
\end{equation}
for transverse shift. Here, $\Delta_+=\Delta_0(k_S+i\chi k_\|\cos\alpha)$, and we define $k_{S}^{\prime}=\sqrt{k_S^2+k_{y}^2}=\sqrt{2m(E_F-U)-k_{z}^2}$.


We have the following observations on these results. First, the basic structure of the expressions follow the discussion
at the end of Sec.~\ref{sec:III}. Namely, for $\varepsilon<|\Delta_+|$, the shift has two contributions, the first is from
the phase of pair potential, and the second is from the evanescent character of mode $\psi_S^+$; for $\varepsilon>|\Delta_+|$, there is only one contribution, from the pairing phase. This feature holds for both longitudinal and transverse shifts.

Second, since the phase of pair potential depends on the chirality $\chi$, the pairing phase contribution in the shift contains the $\chi$ factor. In comparison, the evanescent mode contribution for $\varepsilon<|\Delta_+|$ does not depend on chirality.
For $\varepsilon>|\Delta_+|$, both $\ell_L$ and $\ell_T$ are proportional to $\chi$, so the shift would flip sign if the chirality of the pair potential is reversed.

Third, we note that at $\alpha=0$, i.e., when the incident plane coincides with the $x$-$y$ plane, $\ell_T$ vanishes because the system has a mirror symmetry with respect to the incident plane. Meanwhile, for $\ell_L$, the evanescent mode contribution vanishes due to its $\sin^2\alpha$ dependence, and we find a particularly simple result
\begin{equation}
\ell_L=-\frac{\chi}{k_S},
\end{equation}
which is independent of the excitation energy $\varepsilon$. For $\alpha\neq 0$, $\ell_L$ is an even function of $\alpha$, whereas $\ell_T$ is an odd function.
In Figs.~\ref{figpwave1}(a,b), we plot the variation of the shift components versus $\varepsilon$ at $\alpha\neq 0$. One can see that the
evanescent mode contribution for $\varepsilon<|\Delta_+|$ gives a large contribution close to the superconducting gap.

Finally, regarding the dependence on the incident angle $\gamma$, we note that the pairing phase contribution is an even function of $\gamma$, whereas the evanescent mode contribution is an odd function. It follows that for $\varepsilon>|\Delta_+|$, the curves for $\ell_L$ and $\ell_T$ are symmetric about $\gamma=0$ [see Fig.~\ref{figpwave2}(a,b)]. In comparison, for $\varepsilon<|\Delta_+|$, the curves are generally neither symmetric nor antisymmetric [see Fig.~\ref{figpwave2}(c,d)], due to the presence of both contributions. Interestingly, for $\gamma=0$, i.e., the normal incidence case, we have a simple result: the shift should be along the $y$ direction, with
\begin{equation}\label{30}
  \bm\ell=-\frac{\chi}{\sqrt{2m(E_F-U)}}\hat{y},
\end{equation}
which is proportional to the chirality of pairing and independent of the excitation energy.

\section{d-wave pairing}\label{sec:VI}

\begin{figure}[t]
\includegraphics[width=8.7cm]{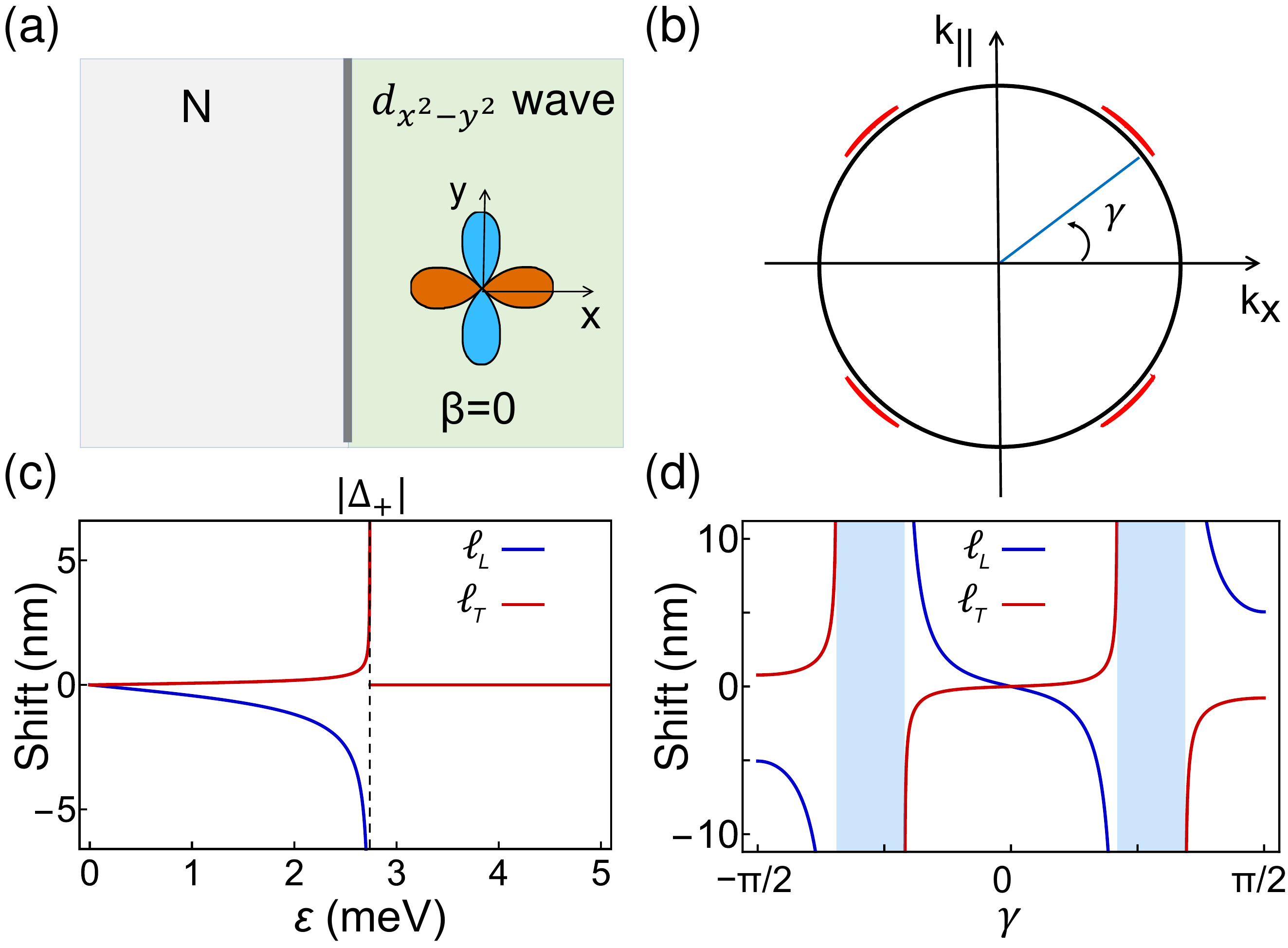} \caption{(a) Illustration of the setup with angle $\beta=0$ for the $d_{x^2-y^2}$-wave case.
(b) illustrates  the equi-energy surfaces of N (black) and S (red) regions at a small excitation energy.
(c) Variation of the shifts versus excitation energy  $\varepsilon$.
(d) Variation of the shifts versus incident angle  $\gamma$.
In the calculation, we take $\Delta_{0}=10\ \rm{meV}\cdot\rm{nm^2}$, $E_{F}=0.1\ {\rm {eV}}$, $m=0.1\  m_e$,
$U=-10\ {\rm {meV}}$, $h=0$ and $\alpha=0.1\pi$. In (c), we take $\gamma=0.05\pi$; In (b) and (d), we take $\varepsilon=1\ {\rm {meV}}.$
\label{figdwave1}}
\end{figure}

The third model we consider is that with a $d_{x^{2}-y^{2}}$-wave pair potential, which we take as {$\boldsymbol\Delta=\Delta_{0}[\cos(2\beta)(k_{x}^{2}-k_{y}^{2})+2\sin(2\beta) k_{x}k_{y}]$}.
Here, $\beta$ is the angle between the $d_{x^{2}-y^{2}}$ wave and the normal direction of the interface ($x$-axis).
For example, the typical cases with $\beta=0$ and $\beta=\pi/4$ are illustrated in Figs.~\ref{figdwave1} and~\ref{figdwave2}. In the following analysis, we again focus on the regime with $E_F\gg \{U, h\}$.

Using Eq.~(\ref{arg3}), we obtain the following analytical expressions for the anomalous shift components:
\begin{eqnarray}
\ell_{L}&=&\frac{2\varepsilon\Delta_{0}\cos\alpha\,\sin (2\beta)(k_{S}^2-k_\|^2)}{k_S\Delta_{+}\sqrt{\Delta_{+}^{2}-\varepsilon^{2}}}\nonumber \\   &&\qquad -\frac{2\varepsilon\Delta_{0}(1+\cos^2\alpha)\cos (2\beta) k_{\|}}{\Delta_{+}\sqrt{\Delta_{+}^{2}-\varepsilon^{2}}}, \label{dGHshift}
\\
\ell_{T}&=&\frac{2\varepsilon \Delta_{0}\sin\alpha \Big[k_{\|}\cos\alpha \cos (2\beta)-k_{S}\sin (2\beta)\Big]}{\Delta_{+}\sqrt{\Delta_{+}^{2}-\varepsilon^{2}}} \label{dIFshift}
\end{eqnarray}
for $\varepsilon< |\Delta_{+}|$, and
\begin{eqnarray}
\ell_{L}=\ell_{T}=0 \label{dIFshift2}
\end{eqnarray}
for $\varepsilon>|\Delta_{+}|$. Here, we have $\Delta_+=\boldsymbol\Delta(k_x=k_S,k_y=k_\|\cos\alpha)$. Typical behaviors of the two shift components versus model parameters have been shown in Figs.~\ref{figdwave1} to \ref{figdwave3}. We have the following observations.

\begin{figure}[t]
\includegraphics[width=8.7cm]{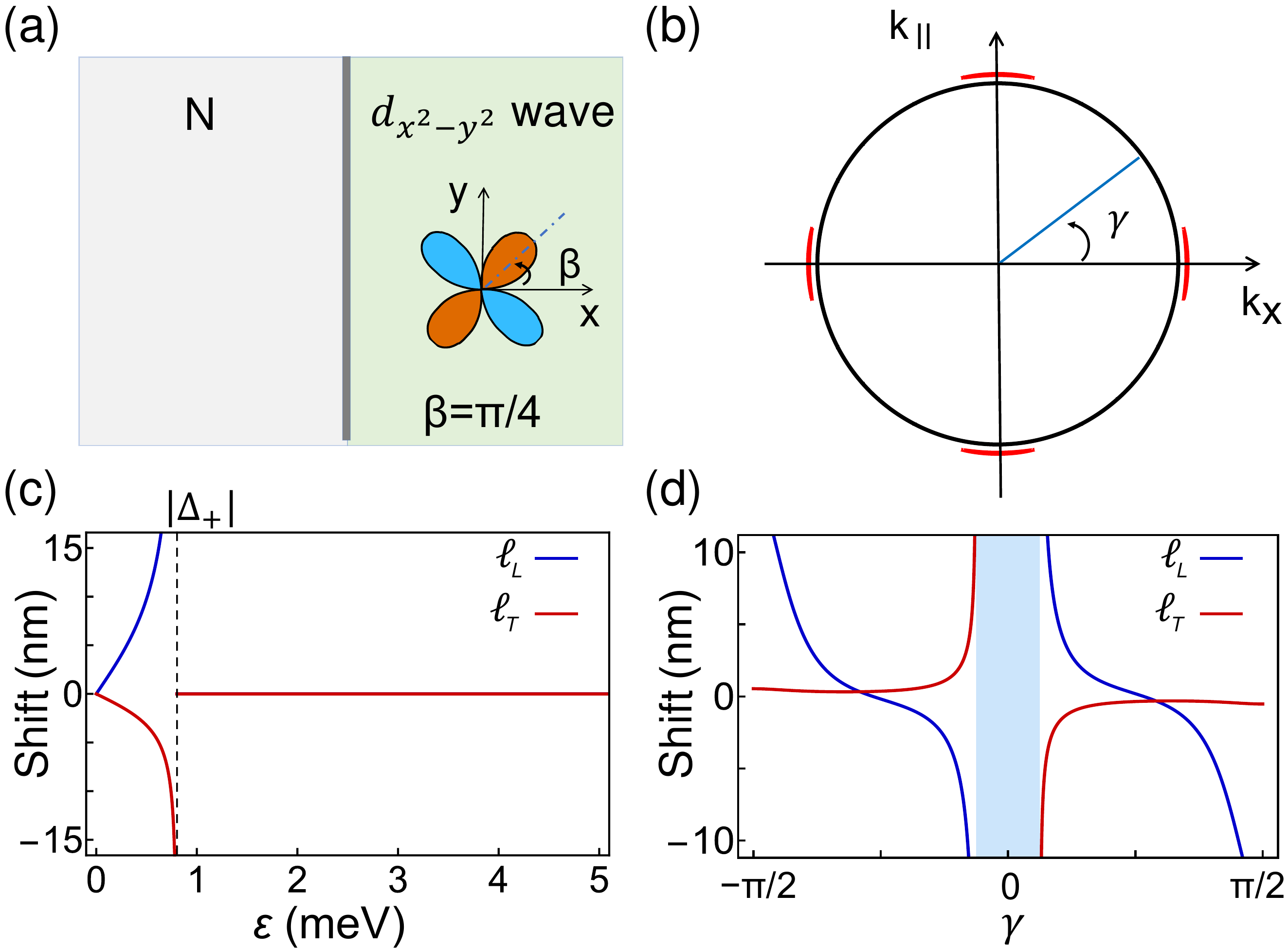} \caption{(a) Illustration of the setup with  angle $\beta=\pi/4$ for the $d_{x^2-y^2}$-wave case.
(b) illustrates  the equi-energy surfaces of N (black) and S (red) regions at a small excitation energy.
(c) Variation of the shifts versus excitation energy  $\varepsilon$.
(d) Variation of the shifts versus incident angle  $\gamma$.
In the figures, except for $\beta$, other parameters are the same as in Fig~\ref{figdwave1}.\label{figdwave2}}
\end{figure}

First, since the pair potential here is real, there is no pairing phase contribution to the shift, and only evanescent mode contribution exists. This explains why the shift vanishes when $\varepsilon>|\Delta_{+}|$, a behavior distinct from the chiral $p$-wave case we discussed in the preceding section. Note that $d_{x^{2}-y^{2}}$-wave pair potential has nodes, where the superconducting gap vanishes. This indicates that for a finite excitation energy $\varepsilon>0$ and large $E_F$, there always exist some range of incident angle in which $\varepsilon>|\Delta_{+}|$ is satisfied and therefore the shift is suppressed.
For example, for the case with $\beta=0$ (see Fig.~\ref{figdwave1}), $|\Delta_{+}|$ vanishes at the two nodes located at $\pm \pi/4$.
This leads to two suppressed zones [marked by the shaded region in Fig.~\ref{figdwave1}(d)] for the incident angle $\gamma$ when it varies from $-\pi /2$ to $\pi/2$. For $\beta\neq 0$, the nodes are shifted, which also shift the locations of the suppressed zones [see Fig.~\ref{figdwave3}(d)]. In the case with $\beta=\pi/4$, there is only one suppressed zone around $\gamma=0$ [see Fig.~\ref{figdwave2}(d)]. By detecting the suppressed zones, one can in principle map out the locations of the nodes.


\begin{figure}[t]
\includegraphics[width=8.7cm]{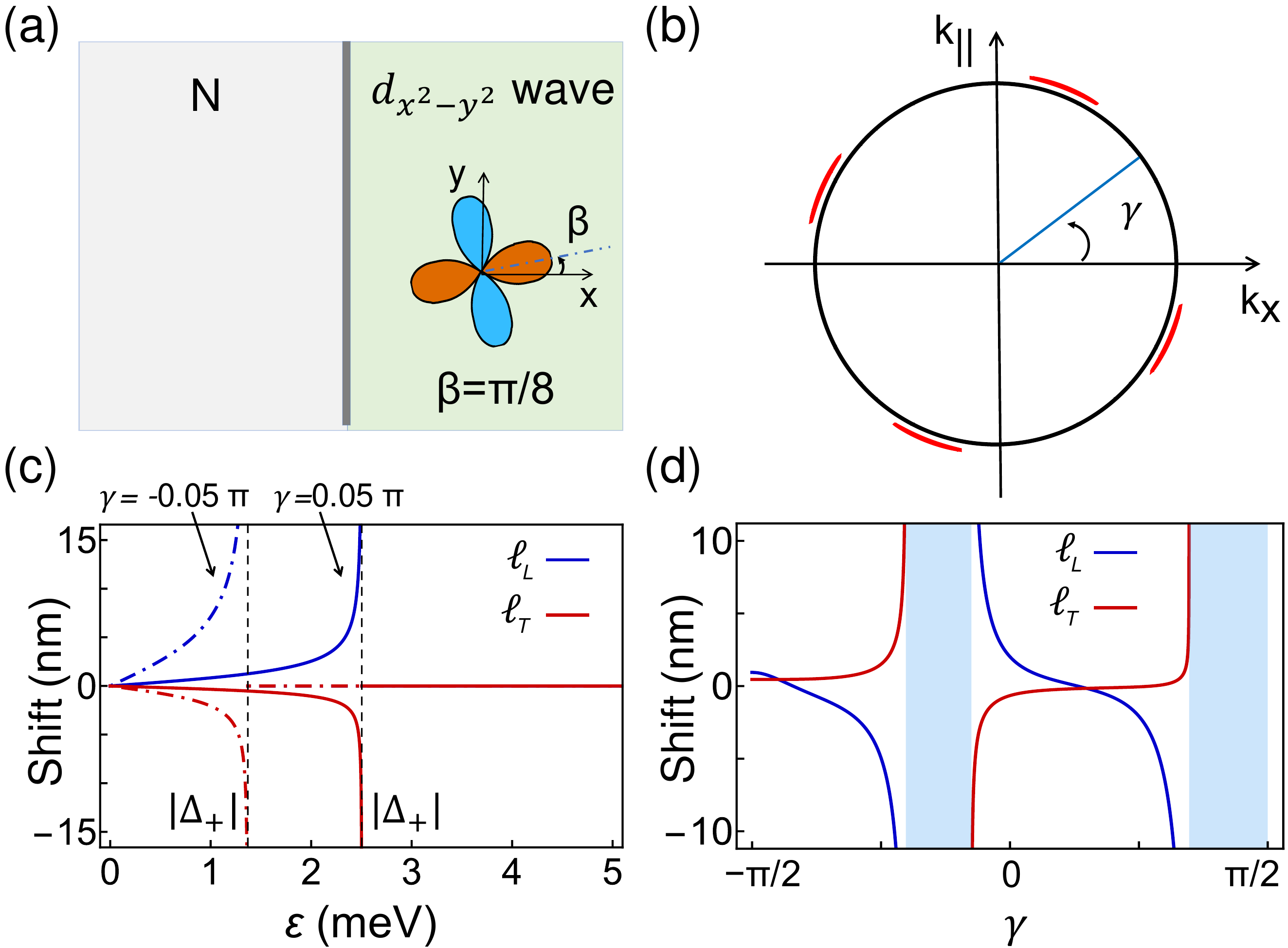} \caption{(a) Illustration of the setup with  angle $\beta=\pi/8$ for the $d_{x^2-y^2}$-wave case.
(b) illustrates  the equi-energy surfaces of N (black) and S (red) regions at a small excitation energy.
(c) Variation of the shifts versus excitation energy  $\varepsilon$.
(d) Variation of the shifts versus incident angle  $\gamma$.
In the figures, except for $\beta$, other parameters are the same as in Fig.~\ref{figdwave1}.\label{figdwave3}}
\end{figure}
Second, regarding the dependence on the angle $\alpha$ of incident plane, from Eqs.~(\ref{dGHshift}-\ref{dIFshift}), we see that $\ell_L$ is an even function, whereas $\ell_T$ is an odd function. This is similar to the chiral $p$-wave case in Sec.~\ref{sec:V}.
It follows that the transverse shift vanishes when $\alpha=0$. Nevertheless, it can be sizable for $\alpha\neq 0$ when the excitation energy is just below the gap $|\Delta_+|$ [see e.g., Fig.~\ref{figdwave3}(c)].


Third, when $\beta=0$, the expressions for the two shift components are simplified as
\begin{equation}
  \ell_{L}= -\frac{\varepsilon\Delta_{0}(3+2\cos 2\alpha) }{\Delta_{+}\sqrt{\Delta_{+}^{2}-\varepsilon^{2}}}k_{\|},
\end{equation}
\begin{equation}
  \ell_{T}=\frac{\varepsilon \Delta_{0} \sin 2\alpha }{\Delta_{+}\sqrt{\Delta_{+}^{2}-\varepsilon^{2}}}k_{\|},
\end{equation}
with $\Delta_+=\Delta_0(k_S^2-k_\|^2\cos^2\alpha)$.
They have opposite signs, and they are both odd functions of the incident angle $\gamma$, as $k_\|$ is odd in $\gamma$ and $\Delta_+$ is even in $\gamma$.
These behaviors are illustrated in Fig.~\ref{figdwave1}(d). For another special case with $\beta=\pi/4$, we have the following simplified expressions:
\begin{equation}
  \ell_{L}=\frac{2\varepsilon\Delta_{0}\cos\alpha(k_{S}^2-k_\|^2)}{k_S\Delta_{+}\sqrt{\Delta_{+}^{2}-\varepsilon^{2}}},
\end{equation}
\begin{equation}
  \ell_{T}=-\frac{2\varepsilon \Delta_{0} \sin\alpha\, k_{S}}{\Delta_{+}\sqrt{\Delta_{+}^{2}-\varepsilon^{2}}},
\end{equation}
with $\Delta_+=2\Delta_0 k_S k_\|\cos\alpha$. One notes that in this case, again, the two components are odd functions of the incident angle $\gamma$ [see Fig.~\ref{figdwave2}(d)], because $\Delta_+$ is odd in $\gamma$ (through $k_\|$). For a general angle $\beta$,
the shift does not have a definite parity with respect to $\gamma$, as shown in Fig.~\ref{figdwave3}(d).

Finally, for normal incidence (with $\gamma=0$), the formula is reduced to
\begin{equation}
  \bm\ell=\frac{2 \varepsilon \Delta_0  k_S \sin 2\beta}{\Delta_{+} \sqrt{\Delta_{+}^2-\varepsilon^2}}\Theta(|\Delta_{+}|-\varepsilon)\hat{y},
\end{equation}
where $\Theta$ is the Heaviside step function, and $\Delta_+=2m(E_F-U)\Delta_0\cos 2\beta$.
This shift is along the $y$ direction and can be nonzero for the general case when $\beta$ is not an integer multiple of $\pi/4$.

\section{Discussion and Conclusion}\label{sec:VII}
{
\begin{figure}[t]
\includegraphics[width=8.7cm]{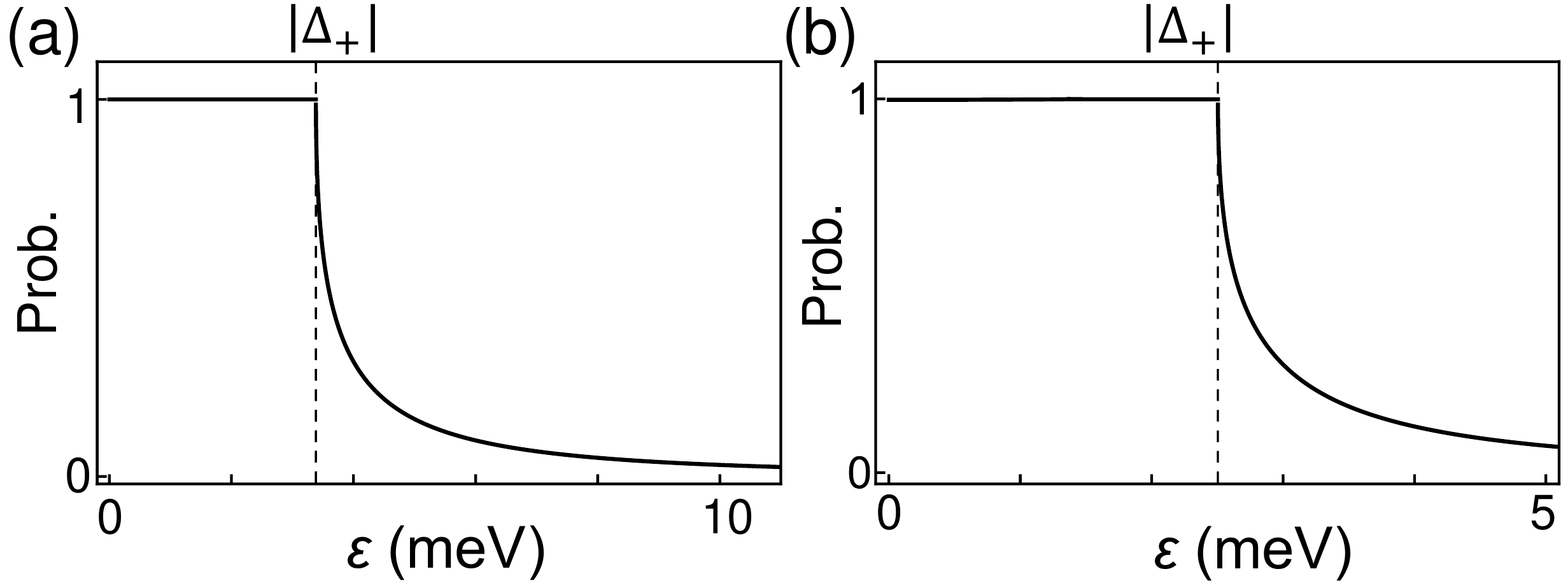} \caption{Variation of Andreev reflection probability $|r_h|^2$ versus excitation energy $\varepsilon$ for (a) chiral $p$-wave case and (b) $d_{x^2-y^2}$ wave case.
In (a), we set the same parameters as Fig.~\ref{figpwave1} (a). And in (b), we set the same parameters as Fig.~\ref{figdwave3} (c) with $\gamma=0.05\pi$.
\label{prob}}
\end{figure}

The anomalous spatial shift in Andreev reflection is connected with the phase of the reflection amplitude $r_h$. The magnitude of $r_h$, on the other hand, determines the probability of Andreev reflection. For large $E_F$, this probability is typically close to unity  when $\varepsilon$ is below the gap $\Delta_+$ and decays with $\varepsilon$ when $\varepsilon$ is above the gap~\cite{Blonder1982}.
For example, the variation of probability $|r_h|^2$ versus $\varepsilon$ is plotted in Fig.~\ref{prob}. for the two cases with chiral $p$-wave pairing and with $d_{x^2-y^2}$-wave pairing, which confirms the above point.

In our model, we used an isotropic Fermi surface. To investigate the effects of Fermi surface anisotropy, we change the normal state Hamiltonian $H_0$ for the S region to be $H_0=-\frac{1}{2m}(\partial_x^2+\partial_y^2)$, which is dispersionless along the $z$ direction. This gives a cylindrical Fermi surface for the S side, representing a very strong anisotropy. For this modified model, analytic results are too complicated to analyze. Nevertheless, we may proceed numerically. As shown in Fig.~\ref{figiso1}(a,b), we find that for small  angle $\alpha$ of the incident plane, the results for the modified model with cylindrical Fermi surface agree very well with those for spherical Fermi surface. This can be understood by noting that when $\alpha\ll 1$, the scattering involves only the states around the Fermi circle with $k_z=0$ in S, which makes little difference between cylindrical and spherical Fermi surfaces. With increased $\alpha$, the results for the cylindrical Fermi surface model does show quantitative difference from the spherical model, but the qualitative features are maintained.
For example, in Fig.~\ref{figiso1}(c,d), we compare the results of the two models for $\alpha=\pi/4$ in the case of $d_{x^2-y^2}$-wave pairing. We see that the two models share similar features, such as existence of suppressed zones, parity with respect to incident angle, and trend of variation against other parameters.
\begin{figure}[t]
\includegraphics[width=8.7cm]{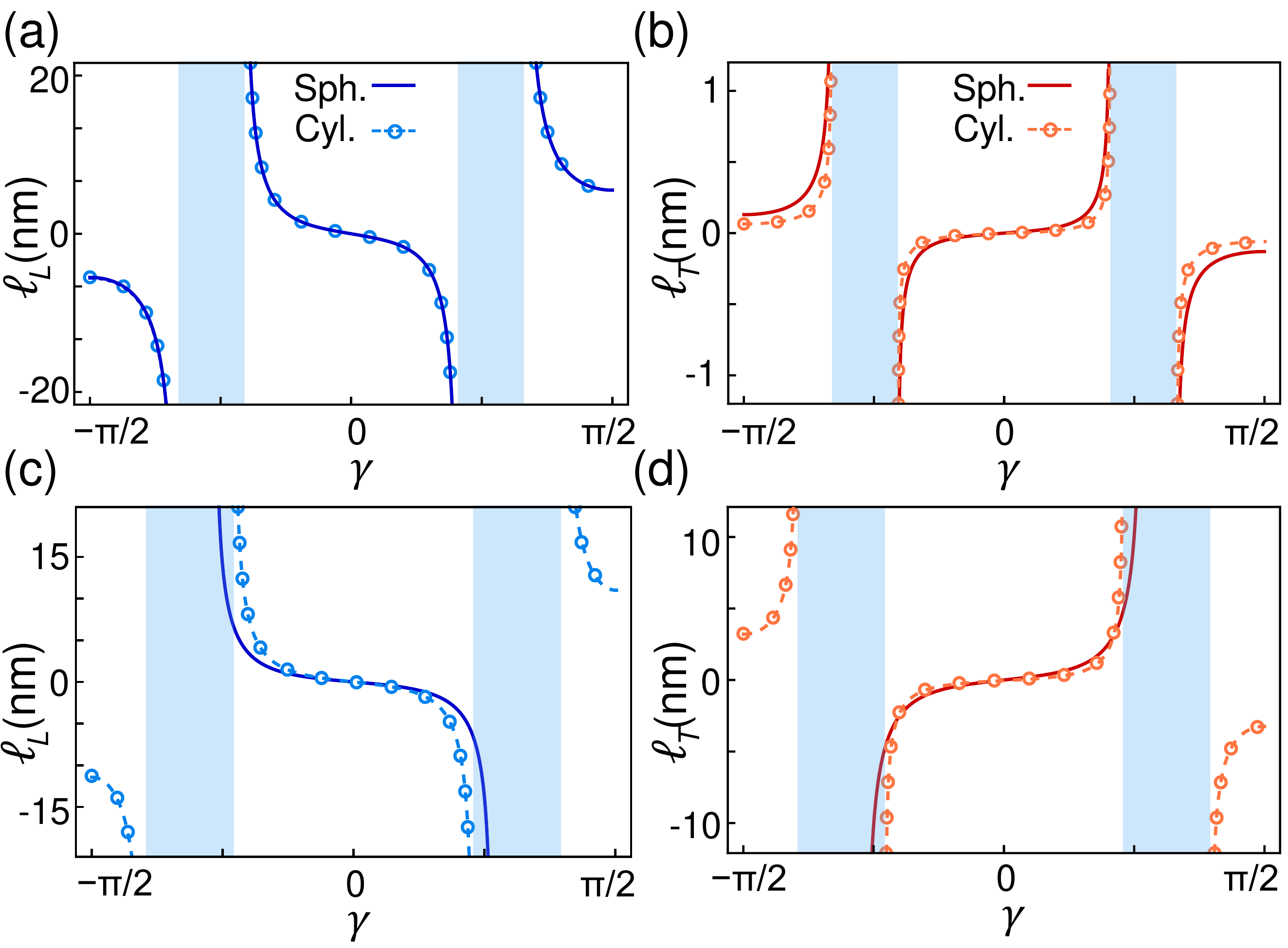} \caption{Results for model with a cylindrical Fermi surface. Here, we consider $d_{x^2-y^2}$-wave case. (a) and (b) are for longitudinal and transverse shift components, respectively, at a small angle
$\alpha=0.01\pi$. In each figure, the lines marked by circles are calculated for the model with cylindrical Fermi surface, and the solid curves are obtained from the formulas in Eqs.~(\ref{dGHshift}) and (\ref{dIFshift}). (c) and (d) are corresponding plots for a larger angle $\alpha=\pi/4$. In these plots, except for $\alpha$, other parameters are the same as in Fig.~\ref{figdwave1}.
\label{figiso1}}
\end{figure}

Regarding experimental detection, the most straightforward way is to prepare a collimated electron beam and let it hit the surface of the superconductor (N side is vacuum in this case), just like the setup of the electron microscope. Then, we detect the Andreev reflected  hole beam. By comparing the trajectories of the incident and the reflected beams, one can extract the shift at the surface during scattering. Our estimation here shows that the shift can reach the magnitude of tens of nanometers, which should be detectable with current technology.  In a NS junction formed by a conventional metal and a superconductor, the shift at the interface may lead to a voltage signal in the transverse direction on the N side close to the interface, when an electric current is driven through the junction. For example, for the junction with a chiral $p$-wave superconductor with interface coinciding with the $y$-$z$ plane, according to Eq.~(\ref{30}), the shift should lead to a voltage drop in the $y$ direction and its sign is determined by the chirality of the pair potential. In addition, the shift may be accumulated by designing heterostructures in which an electron beam can undergo multiple scattering~\cite{Beenakker2009,1Yliu2018}.

In conclusion, we have investigated the anomalous shift in Andreev reflection at an NS interface in the side incidence configuration. The results show rich and distinct behaviors for different types of pairing. For chiral $p$ wave pairing, there are two contributions. The pairing phase contribution which is proportional to chirality and the evanescent mode contribution which is independent of chirality. For excitation energy above the pairing gap, the evanescent mode contribution vanishes, whereas the pairing phase contribution persists, leading to a particularly simple result. For $d_{x^2-y^2}$-wave pairing, only the evanescent mode contribution exists, so the shift vanishes for excitation energy above the gap. Around the nodes of the superconducting gap, there are suppressed zones where the shift vanishes. This offers a way to map out the superconducting nodes. The dependence of the shift on excitation energy, incident angle, and other system parameters are analyzed. These findings deepen our understanding of the fundamental scattering process at the NS interface, offer new methods to characterize superconductors, and may be useful for designing novel superconducting devices.


%
\bibliography{cite}
\bibliographystyle{apsrev4-2}

\end{document}